\documentclass[11pt]{revtex4}
\usepackage{amsmath}
\usepackage{amssymb,epsf}
\usepackage{latexsym}
\usepackage{epsfig}
\usepackage{epstopdf}

\begin{document}

\title{$P$--$V$ criticality of charged dilatonic black holes}
\author{M. H. Dehghani$^{1,2}$\thanks{%
mhd@shirazu.ac.ir}, S. Kamrani $^{1}$ and A. Sheykhi$^{1,2}$\thanks{%
asheykhi@shirazu.ac.ir}}

\begin{abstract}
In this paper, we investigate the critical behavior of charged black holes
of Einstein-Maxwell-dilaton gravity in the presence of two Liouville-type potentials
which make the solution asymptotically neither flat nor AdS and has a parameter $%
\Lambda $ treated as a thermodynamic quantity that can vary. We obtain a
Smarr-type relation for charged dilatonic black holes and find out that the
volume is different from the geometrical volume. We study the analogy of the
Van der Waals liquid-gas system with the charged dilatonic black hole system
while we treat the black hole charge as a fixed external parameter.
Moreover, we show that the critical values for pressure, temperature and
volume are physical provided the coupling constant of dilaton gravity is
less than one and the horizon is sphere. Finally, we calculate the critical
exponents and show that they are universal and are independent of the
details of the system although the thermodynamic quantities depend on the
dilaton parameter and the dimension of the spacetime.
\end{abstract}

\maketitle

\address{$^1$ Physics Department and Biruni Observatory, College of
Sciences, Shiraz University, Shiraz 71454, Iran\\
$^2$ Research Institute for Astronomy and Astrophysics of Maragha
(RIAAM), P.O. Box 55134-441, Maragha, Iran}

\section{Introduction}

In general, the cosmological constant $\Lambda $ is treated as a fixed
parameter when one considers the thermodynamic properties of a black hole.
But based on some basic theories, which state that physical constants are
not fixed a priori and may arise as vacuum expectation values, the
cosmological constant may be treated as a variable parameter. Variation of
the cosmological constant affects the thermodynamic behaviours of
gravitational systems and extends the phase space of the system \cite%
{1,2,3,4,5}. Indeed, thermodynamic properties of black holes in anti
de Sitter (AdS) space are improved in an extended phase space in which the
cosmological constant and its conjugate variable are considered as
thermodynamic pressure and volume, respectively. For instance, the study of
thermodynamic properties of charged AdS black holes shows that a first-order
phase transition between large and small black holes occurs which is
analogous to the Van der Waals liquid-gas phase transition \cite{6}. This
analogy can be improved by extending the thermodynamic phase space in which
the same physical quantities are compared with the liquid-gas system \cite{8}%
. For instance, in nonextended phase space we find that the coexistence
line is in the $\beta -Q$ plane while in the extended phase space, as in the
liquid-gas system, the transition occurs in the $P-T$ plane. The analogy
between a charged black hole and Van der Waals system has been generalized
to the higher dimensional charged black holes, rotating black holes, charged
rotating black holes and Born-Infeld black holes in AdS space \cite{9, 9a}.
Also some relevant discussions in the extended phase space can be found in
Refs. \cite{10}, while this subject in Lovelock and $f(R)$ gravity have been
investigated in Ref. \cite{11}.

In this paper we investigate the critical behavior of ($n+1$)-dimensional
topological charged dilatonic black holes in the extended phase space with
fixed charge parameter (canonical ensemble). The word topological is used
for the black holes whose horizons have topology other than sphere such as
toroid or hyperbola. The presence of the dilaton field in Einstein-Maxwell
theory changes the causal structure of the spacetime and affects on the
thermodynamic properties of the black holes. Due to this fact, it is worth
investigating the effects of dilaton on the the analogy of dilatonic charged
black hole as a Van der Waals system. In an extended phase space, the mass
of black holes does not determine the internal energy but it is related to
the enthalpy, which includes a contribution from the energy formation of the
system. The thermodynamic volume $V$ can then be deduced in terms of the
variables of the black hole spacetime in question which is not necessarily a
geometrical volume defined in the spacetime. This occurs for rotating black
holes \cite{3}, Taub-NUT and Taub-Bolt spacetimes and especially for the
extreme Taub-NUT black hole which has a thermodynamic volume with no
geometrical candidate at all \cite{Nut}. Similar to these cases, we find
that the volume for the cases of dilatonic charged black holes depends on
the coupling constant of dilaton and electromagnetic fields and it is
different from the geometrical volume defined in the spacetime.

The outline of this paper is as follows: In Sec. \ref{sec2}, we review the
thermodynamics of ($n+1$)-dimensional topological black hole solutions of
Einstein-Maxwell-dilaton theory with two Liouville-type potentials. In Sec. %
\ref{Crit}, obtaining the thermodynamic volume and its conjugate quantity
and using the equation of state for charged dilatonic black holes, we
investigate the critical behavior of the system and compare it with those of
a Van der Waals fluid. Also we study the behavior of the system near the
critical point and find the critical exponents. We finish our paper with
some concluding remarks.

\section{Topological Charged dilatonic black holes \label{sec2}}

In this section, we review the thermodynamics of $(n+1)$-dimensional
topological charged black holes in dilaton gravity. The action of
Einstein-Maxwell theory in the presence of a dilaton field is \cite{17}:%
\begin{equation}
I=\frac{1}{16\pi }\int d^{n+1}x\sqrt{-g}\left( \mathcal{R}\text{ }-\frac{4}{%
n-1}(\nabla \Phi )^{2}-V(\Phi )-e^{-4\alpha \Phi /(n-1)}F_{\mu \nu }F^{\mu
\nu }\right)  \label{Act1}
\end{equation}%
where $\Phi $ is the dilaton field, $F_{\mu \nu }$ is the Maxwell field
strength defined as $F_{\mu \nu }=\partial _{\lbrack \mu }A_{\nu ]}$ with
vector potential $A$ and $\alpha $ is the the coupling constant of the
scalar and electromagnetic fields. In order to have topological dilatonic black holes,
$V(\Phi )$ in Eq. (\ref{Act1}) should be chosen as \cite{18}
\begin{equation}
V(\Phi ) =2\Lambda e^{4\alpha \Phi /(n-1)}+\frac{k(n-1)(n-2)\alpha ^{2}}{%
b^{2}(\alpha ^{2}-1)} e^{4\Phi /[(n-1)\alpha ]},  \label{VPhi}
\end{equation}%
where $\Lambda $ is a parameter which will be treated as a thermodynamic
quantity.

The action (\ref{Act1}) with potential (\ref{VPhi}) admits a static black
hole solution with metric \cite{18}
\begin{equation}
ds^{2}=-f(r)dt^{2}+\frac{dr^{2}}{f(r)}+r^{2}R(r)^{2}d\Omega _{k}^{2},
\label{metric}
\end{equation}%
where $d\Omega _{k}^{2}$ stands for the line element of an $(n-1)$%
-dimensional maximal symmetric space with constant curvature $(n-1)(n-2)k$.
We denote the area of this $(n-1)$-dimensional surface with unit radius by $%
\omega _{n-1}$.\ Considering the field equations of
Einstein--Maxwell--dilaton gravity \cite{18} with metric (\ref{metric}) and
using $R(r)=e^{2\alpha \Phi /(n-1)}$, one can show that the metric function $%
f(r)$, the dilaton field $\Phi (r)$ and the electromagnetic field $F_{\mu
\nu }$\ are given as \cite{18}
\begin{eqnarray}
f(r) &=&\frac{2\Lambda \left( \alpha ^{2}+1\right) ^{2}b^{2\gamma }}{%
(n-1)\left( \alpha ^{2}-n\right) }r^{2(1-\gamma )}-\frac{k\left( n-2\right)
\left( \alpha ^{2}+1\right) ^{2}{b}^{-2\gamma }{r}^{2\gamma }}{\left( \alpha
^{2}-1\right) \left( \alpha ^{2}+n-2\right) }  \notag \\
&&-\frac{m}{r^{\left( n-1\right) \left( 1-\gamma \right) -1}}+\frac{%
2q^{2}(\alpha ^{2}+1)^{2}b^{-2(n-2)\gamma }}{(n-1)(\alpha ^{2}+n-2)}%
r^{2(n-2)(\gamma -1)}
\end{eqnarray}

\begin{equation}
\Phi (r)=\frac{(n-1)\alpha }{2(1+\alpha ^{2})}\ln (\frac{b}{r}),\text{ \ \ \
\ }F_{tr}=\frac{qe^{4\alpha \Phi /(n-1)}}{(rR)^{n-1}}  \label{Ftr}
\end{equation}%
where $b$ is a nonzero positive arbitrary constant, $\gamma =\alpha
^{2}/\left( \alpha ^{2}+1\right) $ and the parameters $m$ and $q$ are mass
and charge parameters, respectively. The mass parameter $m$ can be written
in term of the horizon radius as \cite{18} \newline
\begin{eqnarray}
m(r_{+}) &=&\frac{2\Lambda \left( \alpha ^{2}+1\right) ^{2}b^{2\gamma }}{%
(n-1)\left( \alpha ^{2}-n\right) }r_{+}^{n(1-\gamma )-\gamma }-\frac{%
k(n-2)\left( \alpha ^{2}+1\right) ^{2}b^{-2\gamma }}{\left( \alpha
^{2}-1\right) (n+\alpha ^{2}-2)}r_{+}^{n-2+\gamma (3-n)}  \notag \\
&&+\frac{2q^{2}\left( \alpha ^{2}+1\right) ^{2}b^{-2\gamma (n-2)}}{%
(n-1)(n+\alpha ^{2}-2)}r_{+}^{(n-3)(\gamma -1)-1},
\end{eqnarray}%
where $r_{+}$ denotes the radius of the event horizon which is the largest
root of $f(r_{+})=0$. The ADM mass and the electric charge of the black hole
are \cite{18} \newline
\begin{equation}
M=\frac{b^{(n-1)\gamma }(n-1)\omega _{n-1}}{16\pi \left( \alpha
^{2}+1\right) }m,\text{ \ \ }Q=\frac{q\omega _{n-1}}{4\pi }.
\end{equation}%
The Hawking temperature of the topological black hole on outer horizon $%
r_{+} $ can be calculated as \cite{18} \newline
\begin{eqnarray}
\mathcal{T} &=&\frac{f^{\prime }(r_{+})}{4\pi }=-\frac{k(n-2)\left( \alpha
^{2}+1\right) b^{-2\gamma }}{4\pi \left( \alpha ^{2}-1\right) }%
r_{+}^{2\gamma -1}  \notag \\
&&-\frac{\Lambda \left( \alpha ^{2}+1\right) b^{2\gamma }}{2\pi (n-1)}%
r_{+}^{1-2\gamma }-\frac{q^{2}\left( \alpha ^{2}+1\right) b^{-2(n-2)\gamma }%
}{2\pi (n-1)}r_{+}^{(2n-3)(\gamma -1)-\gamma }.  \label{Temp}
\end{eqnarray}

Using the so called area law of the entropy in Einstein gravity which states
that the entropy of the black hole is a quarter of the event horizon area
\cite{20}, one obtains \cite{18}
\begin{equation}
S=\frac{b^{(n-1)\gamma }r_{+}^{(n-1)(1-\gamma )}\omega _{n-1}}{4}
\label{entropy}
\end{equation}%
The electric potential $U$, measured at infinity with respect to the
horizon, is defined by
\begin{equation}
U =A_{\mu }\chi ^{\mu }\left\vert _{r\rightarrow \infty }-A_{\mu }\chi ^{\mu
}\right\vert _{r=r_{+}},
\end{equation}%
where $\chi =\partial _{t}$ is the null generator of the horizon. Since the
gauge potential $A_{t}$ corresponding to the electromagnetic field (\ref{Ftr}%
) can be written as
\begin{eqnarray}
A_{t} &=&\frac{qb^{(3-n)\gamma }}{\Upsilon r^{\Upsilon }},  \notag \\
\Upsilon &=&(n-3)(1-\gamma )+1,  \label{At}
\end{eqnarray}%
the electric potential is \cite{18}
\begin{equation}
U =\frac{qb^{(3-n)\gamma }}{\Upsilon {r_{+}}^{\Upsilon }}.
\end{equation}
It is worth mentioning that the $k=1$ black holes are thermally unstable.
Indeed, as one can see in Ref. \cite{18}, a Hawking-Page phase transition
can happen for these black holes.

\section{Phase transition of charged dilatonic black holes in ($n+1$%
) dimensions \label{Crit}}

Here, we investigate the thermodynamics of $(n+1)$-dimensional charged
dilatonic black holes in an extended phase space, treating the cosmological
constant and its conjugate quantity as thermodynamic variables associated
with the pressure and volume, respectively. Since we are going to discuss
the thermodynamics of the black hole in the extended phase space by
introducing the pressure proportional to the cosmological constant, the
black hole mass $M$ should be considered as the enthalpy $H\equiv M$ rather
than the internal energy of the gravitational system \cite{2}. Using the
fact that the entropy of black hole is a quarter of the area of the horizon,
the thermodynamic volume $V=\int 4Sdr_{+}$ is obtained as%
\begin{eqnarray}
V &=&\frac{b^{(n-1)\gamma }\omega _{n-1}}{\varepsilon }r_{+}^{\varepsilon },
\label{volume} \\
\varepsilon &=&(n-1)(1-\gamma )+1=\frac{n+\alpha ^{2}}{1+\alpha ^{2}},
\label{eps}
\end{eqnarray}%
which is different from the geometrical volume. Using the first law of
thermodynamics
\begin{equation}
dM=\mathcal{T}dS+U dQ+Vd\mathcal{P},
\end{equation}%
one can show that the pressure, which is the conjugate quantity of the
thermodynamic volume, is
\begin{equation}
\mathcal{P}=-\frac{\left[ n-\gamma (n-1)\right] b^{2\gamma }}{8\pi \left[
n-\gamma (n+1)\right] r_{+}^{2\gamma }}\Lambda =-\frac{\left( n+\alpha
^{2}\right) b^{2\gamma }}{8\pi \left( n-\alpha ^{2}\right) r_{+}^{2\gamma }}%
\Lambda .  \label{press}
\end{equation}%
One may note that the above $\mathcal{P}$ is proportional to the
cosmological constant $\Lambda $ and reduces to the pressure for
Reissner-Nordstrum black hole in the absence of dilaton ($\gamma =0$). Also,
one should note that the pressure is positive provided $\alpha ^{2}<n$ ($%
\gamma <n/(n+1)$). The above thermodynamic quantities satisfy the following
Smarr formula:
\begin{equation}
M=\frac{(n-1)(1-\gamma )}{\Upsilon }\mathcal{T}S+U Q+\frac{(4\gamma -2)}{%
\Upsilon }V\mathcal{P}.
\end{equation}%
Now, we study the analogy of the liquid--gas phase transition of the Van der
Waals fluid with the phase transition of the charged dilatonic black hole
system in the extended phase space in canonical ensemble, in which we treat
the black hole charge $Q$ as a fixed external parameter.

\subsection{Equation of state}

Using Eqs. (\ref{Temp}) and (\ref{press}) for a fixed charge $Q$, one may
obtain the equation of state $\mathcal{P}(V,\mathcal{T})$ as
\begin{equation}
P=\frac{(n+\alpha ^{2})(n-1) T}{4(n-\alpha ^{2})(1+\alpha ^{2})r_{+}}+\frac{%
k(n-2)(n-1)(n+\alpha ^{2})b^{-2\gamma }}{16\pi (\alpha ^{2}-1)(n-\alpha
^{2})r_{+}^{2-2\gamma }}+\frac{(n+\alpha ^{2})q^{2}b^{-2(n-2)\gamma }}{8\pi
(n-\alpha ^{2})r_{+}^{1-(2n-3)(\gamma -1)+\gamma }}  \label{eq of state}
\end{equation}%
where $r_{+}$ is a function of the thermodynamic volume $V$ given in Eq. (%
\ref{volume}). Before we proceed further, we perform the dimensional
analysis to translate the `geometric' equation of state (\ref{eq of state})
to a physical one. The physical pressure and temperature are given by
\begin{equation}
\mathcal{P}=\frac{\hbar c}{l_{p}^{2}} P,\quad \mathcal{T}=\frac{\hbar c}{%
\kappa }T,
\end{equation}%
where the Planck length reads $l_{p}^{2}=\hbar G/c^{3}$ and $\kappa $ is the
Boltzmann constant. Therefore \newline
\begin{equation}
\mathcal{P}=\frac{(n+\alpha ^{2})(n-1)\kappa \mathcal{T}}{%
4l_{p}^{2}(n-\alpha ^{2})(1+\alpha ^{2})r_{+}}+\frac{k(n-2)(n-1)(n+\alpha
^{2})\hbar cb^{-2\gamma }}{l_{p}^{2}16\pi (\alpha ^{2}-1)(n-\alpha
^{2})r_{+}^{2-2\gamma }}+\frac{(n+\alpha ^{2})q^{2}\hbar cb^{-2(n-2)\gamma }%
}{l_{p}^{2}8\pi (n-\alpha ^{2})r_{+}^{1-(2n-3)(\gamma -1)+\gamma }}.
\label{EqstatePh}
\end{equation}%
Comparing with the Van der Waals equation \cite{8}, we conclude that one
should identify the specific volume $v$ of the fluid with the horizon radius
of black hole as
\begin{equation}
v=\frac{4l_{p}^{2}(1+\alpha ^{2})(n-\alpha ^{2})r_{+}}{(n-1)(n+\alpha ^{2})}
\label{volume2}
\end{equation}%
Using the above identification and returning to geometric units, the
equation of state (\ref{eq of state}) can be written as
\begin{eqnarray}
P &=&\frac{T}{v}+\frac{k(n-2)(n-\alpha ^{2})^{1-2\gamma }b^{-2\gamma }}{%
4^{2\gamma }\pi (\alpha ^{2}-1)(\alpha ^{2}+1)^{2\gamma -2}[(n+\alpha
^{2})(n-1)]^{1-2\gamma }v^{2-2\gamma }}  \notag \\
&&+\frac{q^{2}b^{-2(n-2)\gamma }(n+\alpha ^{2})^{(2n-3)(\gamma -1)-\gamma
}(n-1)^{(2n-3)(\gamma -1)-\gamma -1}}{2^{(4n-6)(\gamma -1)-2\gamma +1}\pi
(n-\alpha ^{2})^{(2n-3)(\gamma -1)-\gamma }(1+\alpha ^{2})^{(2n-3)(\gamma
-1)-\gamma -1}v^{\gamma -(2n-3)(\gamma -1)+1}}.  \label{PvT}
\end{eqnarray}%
\begin{figure}[h]
$%
\begin{array}{cc}
\epsfxsize=7cm \epsffile{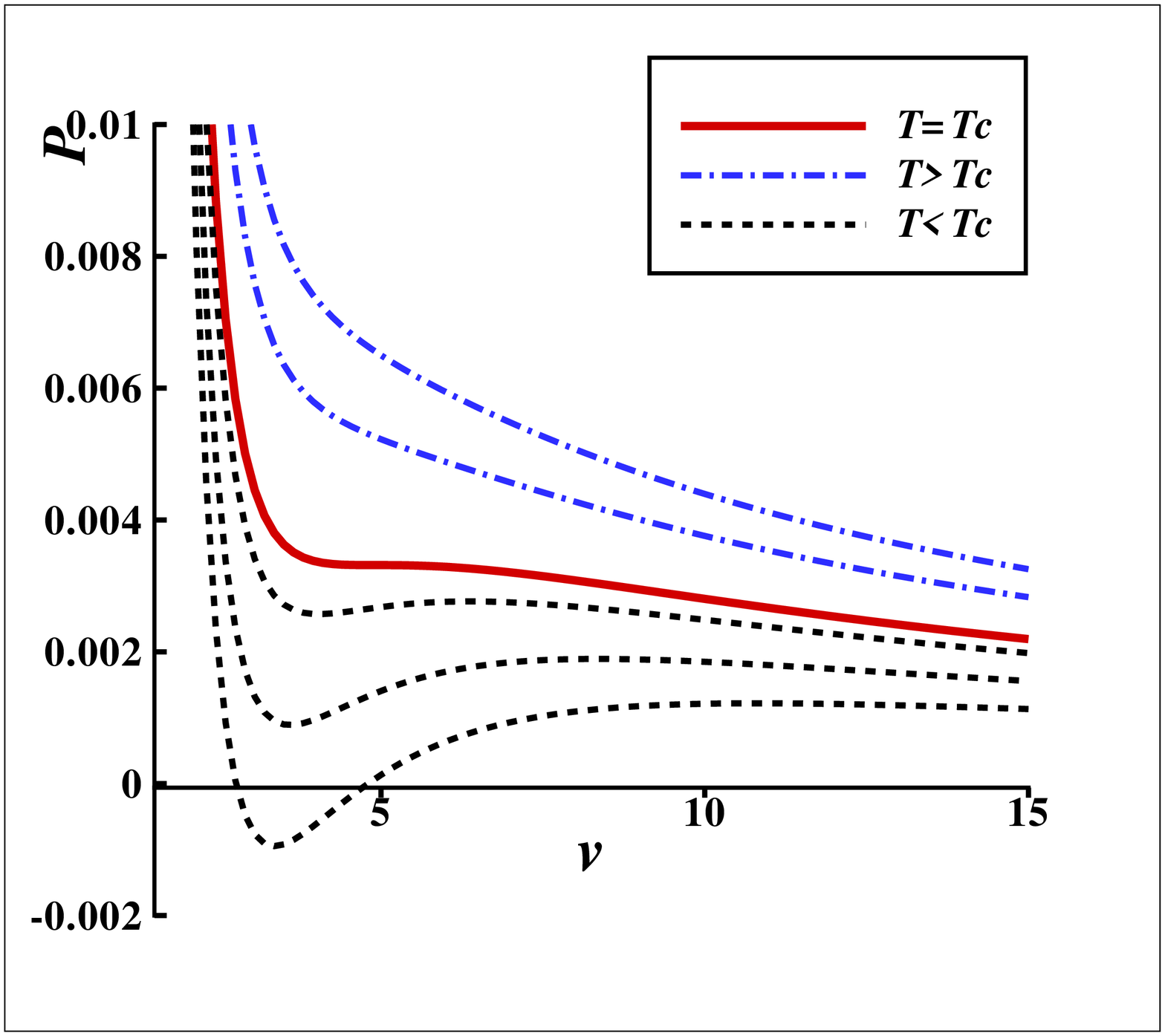} & \epsfxsize=7cm \epsffile{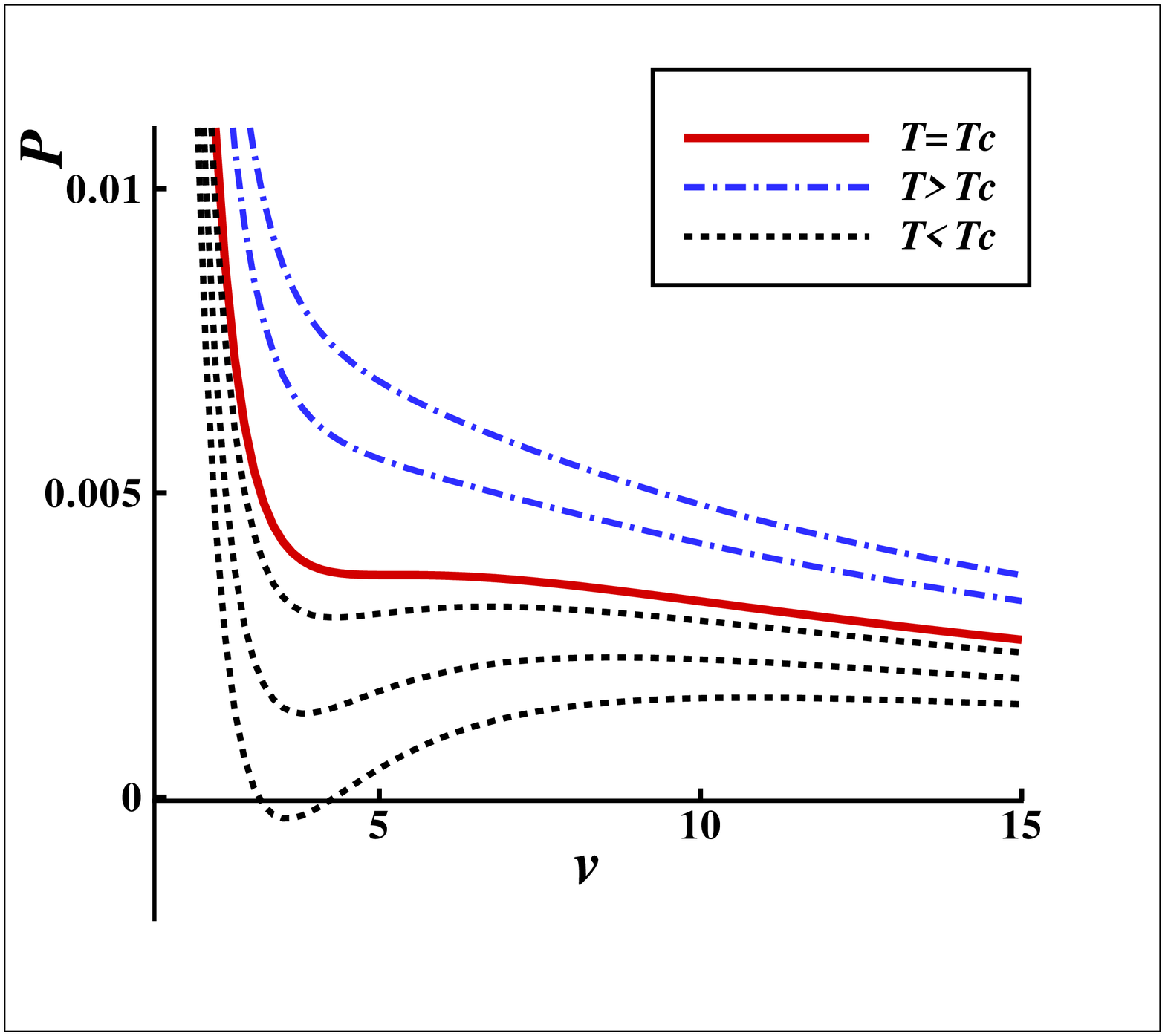}%
\end{array}%
$%
\caption{$P-v$ diagram of charged dilatonic black hole in the case of $k=1$
for $n=3$, $q=b=1.0$, $\protect\alpha =0$ (left) and $\protect\alpha =0.3$
(right).}
\label{Fig1}
\end{figure}
\begin{figure}[h]
$%
\begin{array}{cc}
\epsfxsize=7cm \epsffile{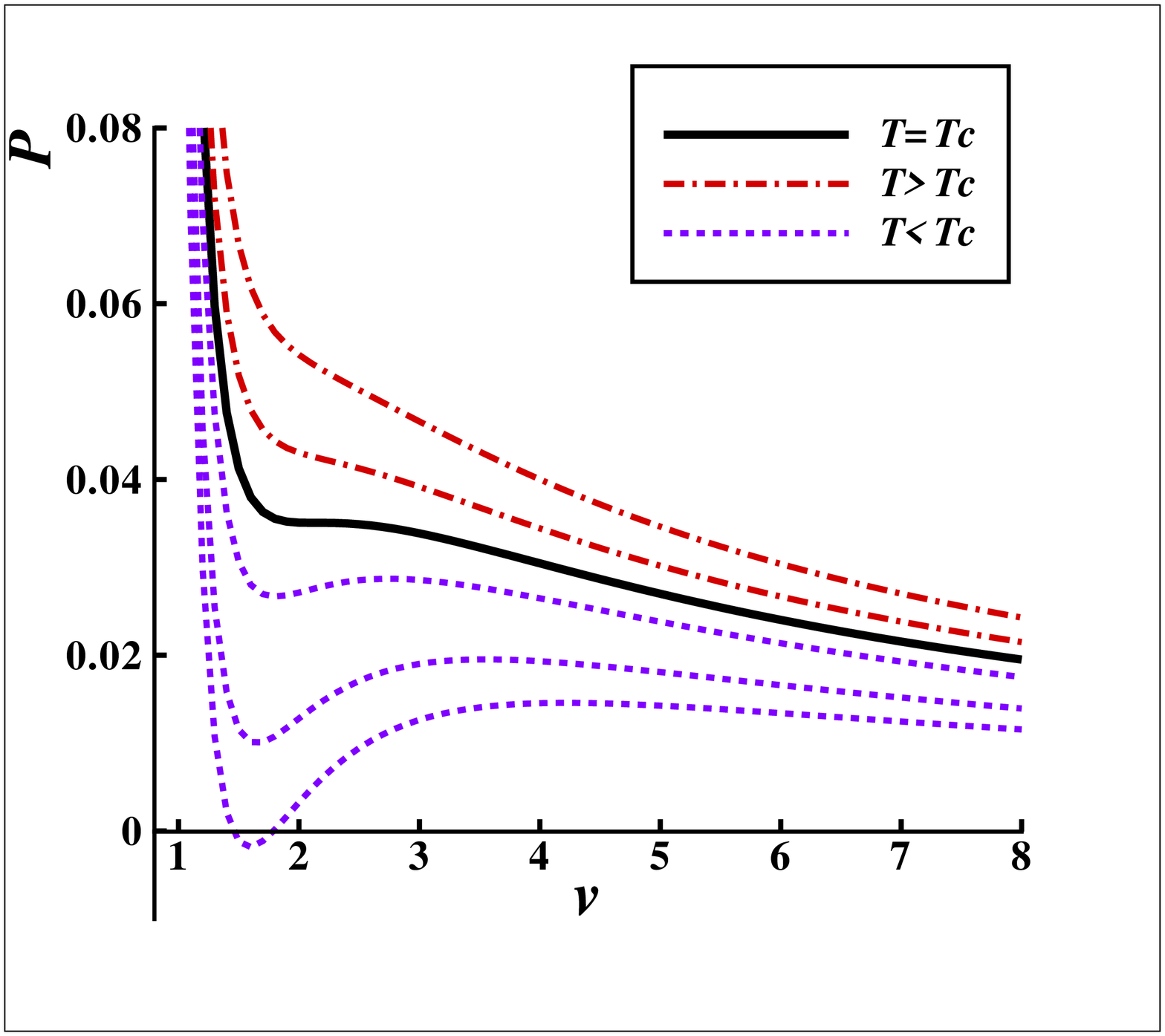} & \epsfxsize=7cm \epsffile{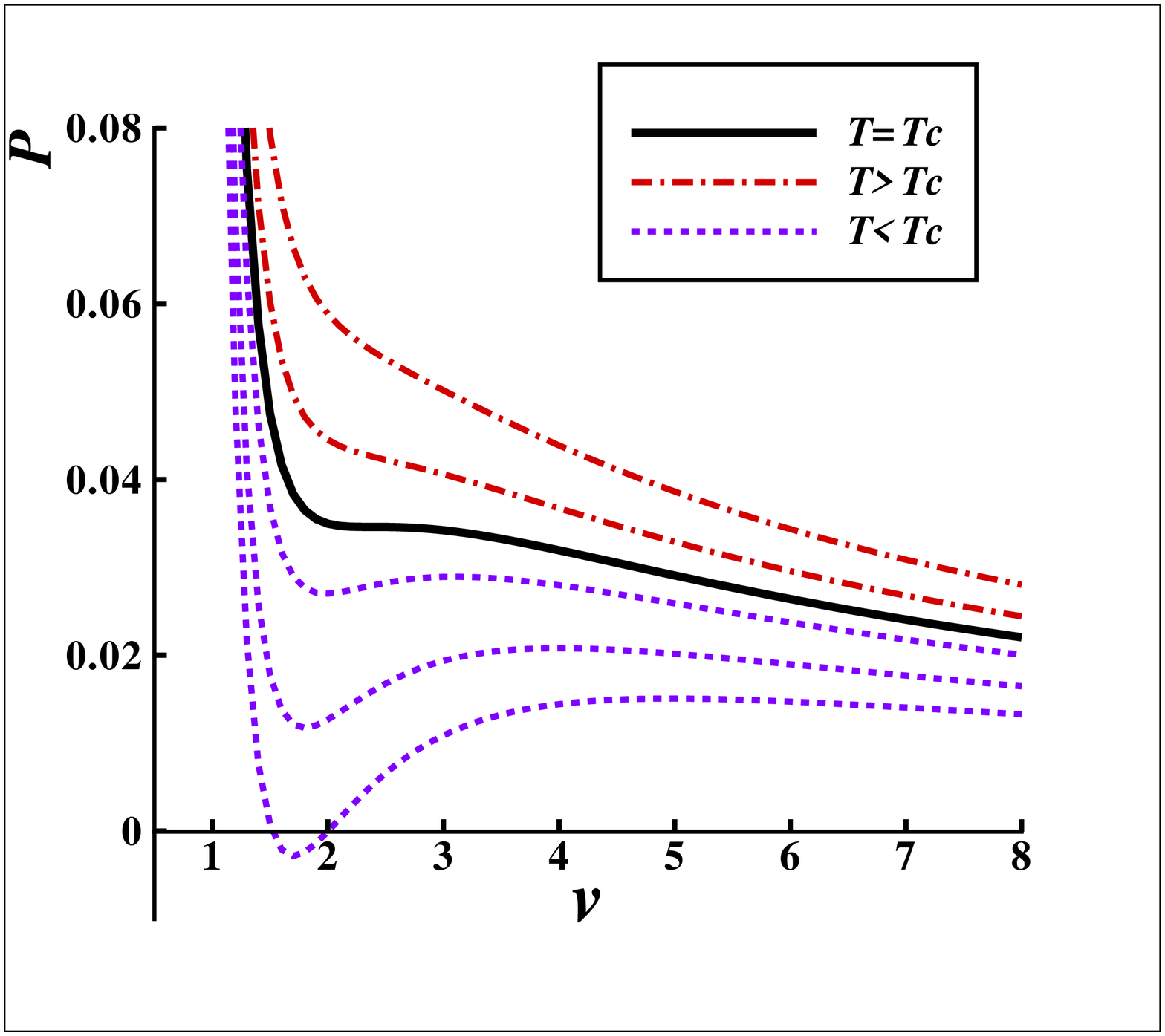}%
\end{array}%
$%
\caption{$P-v$ diagram of charged dilatonic black hole in the case of $k=1$
for $n=4$, $q=b=1.0$, $\protect\alpha =0.3$ (left) and $\protect\alpha =0.5$
(right).}
\label{Fig2}
\end{figure}

Now, we are ready to investigate the possibility of critical behavior for
the case of charged dilatonic black holes. To do this, we calculate the
pressure and volume at the critical point which are known as the critical
pressure and the critical volume. This can be done by solving the following
equations:
\begin{equation}
\frac{\partial P}{\partial v}\Big|_{T_{c}}=0,\quad \frac{\partial ^{2}P}{%
\partial v^{2}}\Big|_{T_{c}}=0.  \label{P1}
\end{equation}%
One obtains
\begin{eqnarray}
v_{c} &=&\left[ \frac{2q^{2}(n-1+\alpha ^{2})(2n-3+\alpha ^{2})b^{6\gamma
-2n\gamma }}{k(n-2)(n-1)}\right] ^{1/2\Upsilon }\times \left[ \frac{%
4(n-\alpha ^{2})(1+\alpha ^{2})}{(n-1)(n+\alpha ^{2})}\right] \\
T_{c} &=&\left[ \frac{(n-1)}{2q^{2}(n-1+\alpha ^{2})}\right] ^{(1-2\gamma
)/2\Upsilon }\times \left[ \frac{(2n-3+\alpha ^{2})}{k(n-2)}\right]
^{[(2n-3)(\gamma -1)-\gamma ]/2\Upsilon }  \notag \\
&&\times \left[ \frac{(\alpha ^{2}+n-2)b^{-\gamma (n-1)/\Upsilon }}{\pi
(1-\alpha ^{2})}\right] \\
P_{c} &=&\left[ \frac{k(n-1)(n-2)}{(2n-3+\alpha ^{2})(2n-2+2\alpha ^{2})}%
\right] ^{[\gamma -(2n-3)(\gamma -1)+1]/2\Upsilon }  \notag \\
&&\times \left[ \frac{(n+\alpha ^{2})(2n-3+\alpha ^{2})(n-2+\alpha
^{2})b^{-2\gamma /\Upsilon }}{8\pi (1+\alpha ^{2})(n-\alpha
^{2})q^{2(1-\gamma )/\Upsilon }}\right]
\end{eqnarray}%
This gives the following universal ratio%
\begin{equation}
\rho _{c}=\frac{P_{c}v_{c}}{T_{c}}=\frac{(1-\alpha ^{2})(2n-3+\alpha ^{2})}{%
4(n-1+\alpha ^{2})}  \label{universal ratio}
\end{equation}%
Note that for $\alpha =0$ in four-dimensional spacetime, we recover the
ratio $\rho _{c}=3/8$ which is the characteristic of Van der Waals fluid. It
is worth mentioning that a positive value for universal ratio $\rho _{c}$ is
guaranteed provided $\alpha <1$. In this case, Eq. (\ref{PvT}) shows that
only for the case of $k=1$, the critical behavior occurs. This is due to the
fact that for the cases of $k=0$ and $k=-1$ with $\alpha <1$, all the terms
in Eq. (\ref{PvT}) are positive and therefore no critical behavior will
occur. The $P-v$ isothermal diagrams of the case $k=1$ for different values
of dilaton coupling constant $\alpha <1$ in $4$ and $5$ dimensions are shown
in Figs. \ref{Fig1} and \ref{Fig2}. As in the case of Van der Waals gas,
there is a critical point which is a point of inflection on the critical
isotherm. Obviously, for $T<T_{c}$ there is a small/large black hole phase
transition in the system.

\subsection{Gibbs free energy}

\begin{figure}[h]
$%
\begin{array}{cc}
\epsfxsize=7cm \epsffile{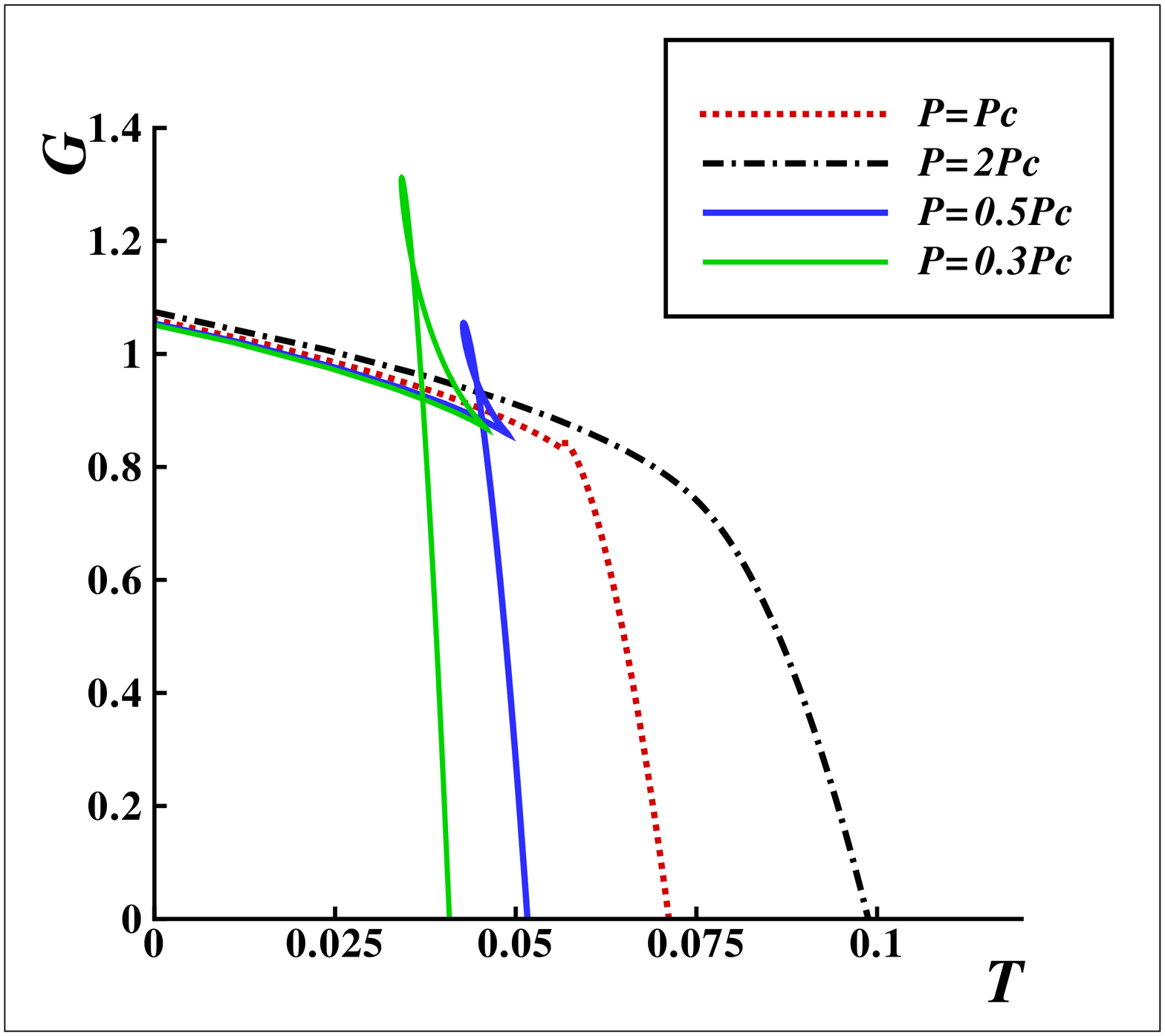} & \epsfxsize=7cm \epsffile{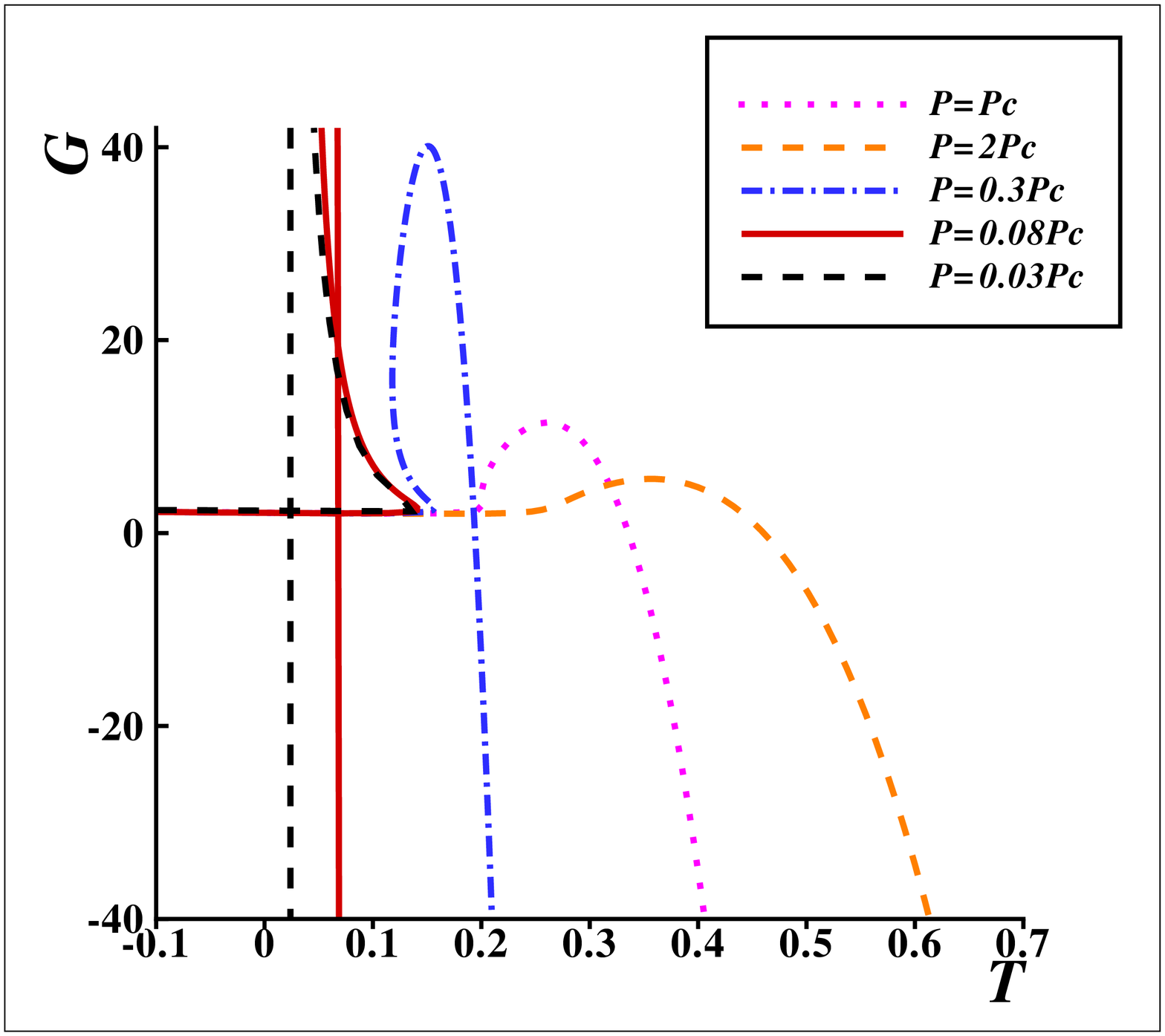}%
\end{array}%
$%
\caption{Gibbs free energy of charged dilatonic black hole with $\protect%
\alpha=0.3$, $q=b=1.0$ for $n=3 $ (left) and $n=4 $ (right).}
\label{Gibbs fig}
\end{figure}

\begin{figure}[h]
$%
\begin{array}{cc}
\epsfxsize=7cm \epsffile{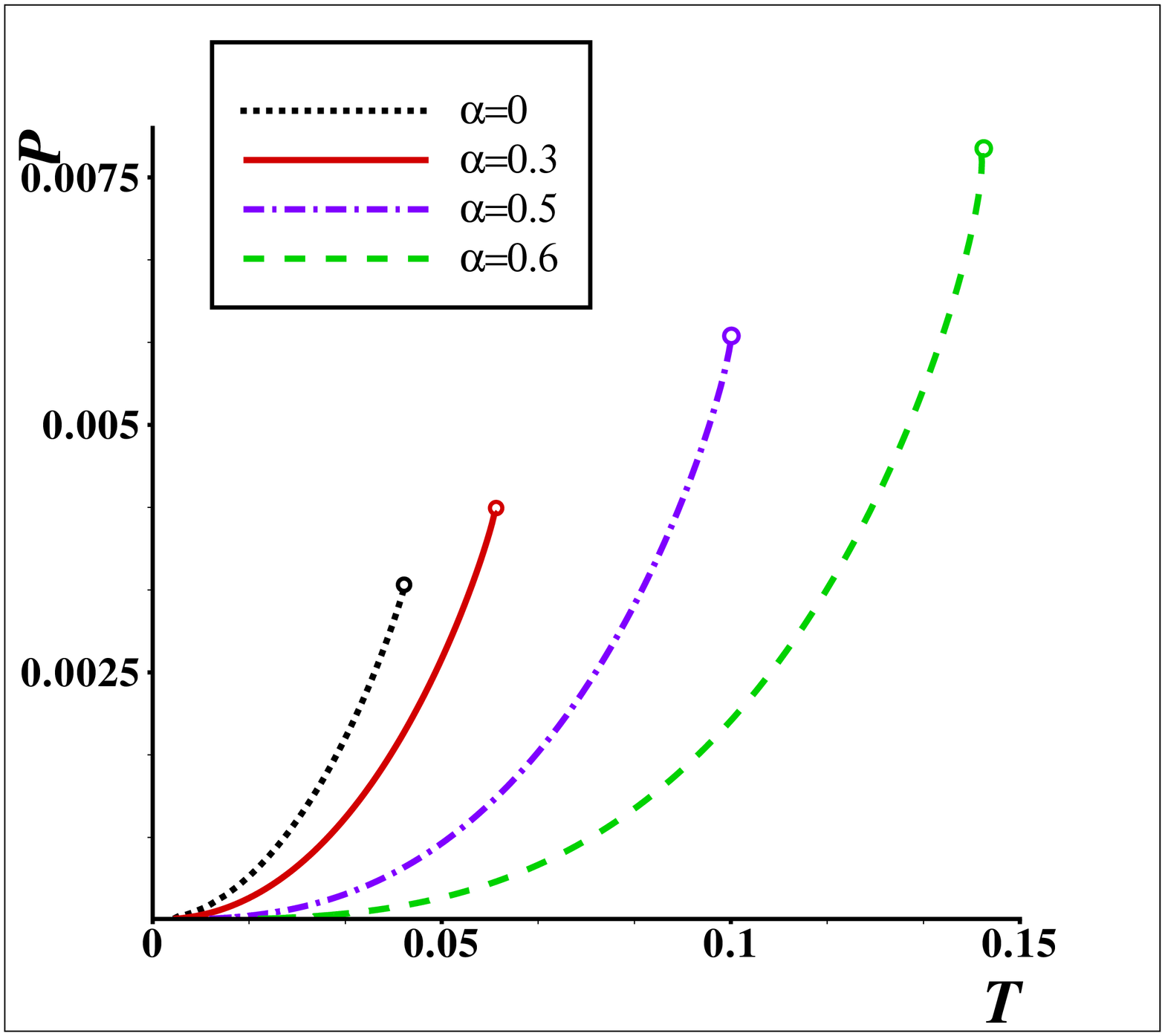} & \epsfxsize=7cm \epsffile{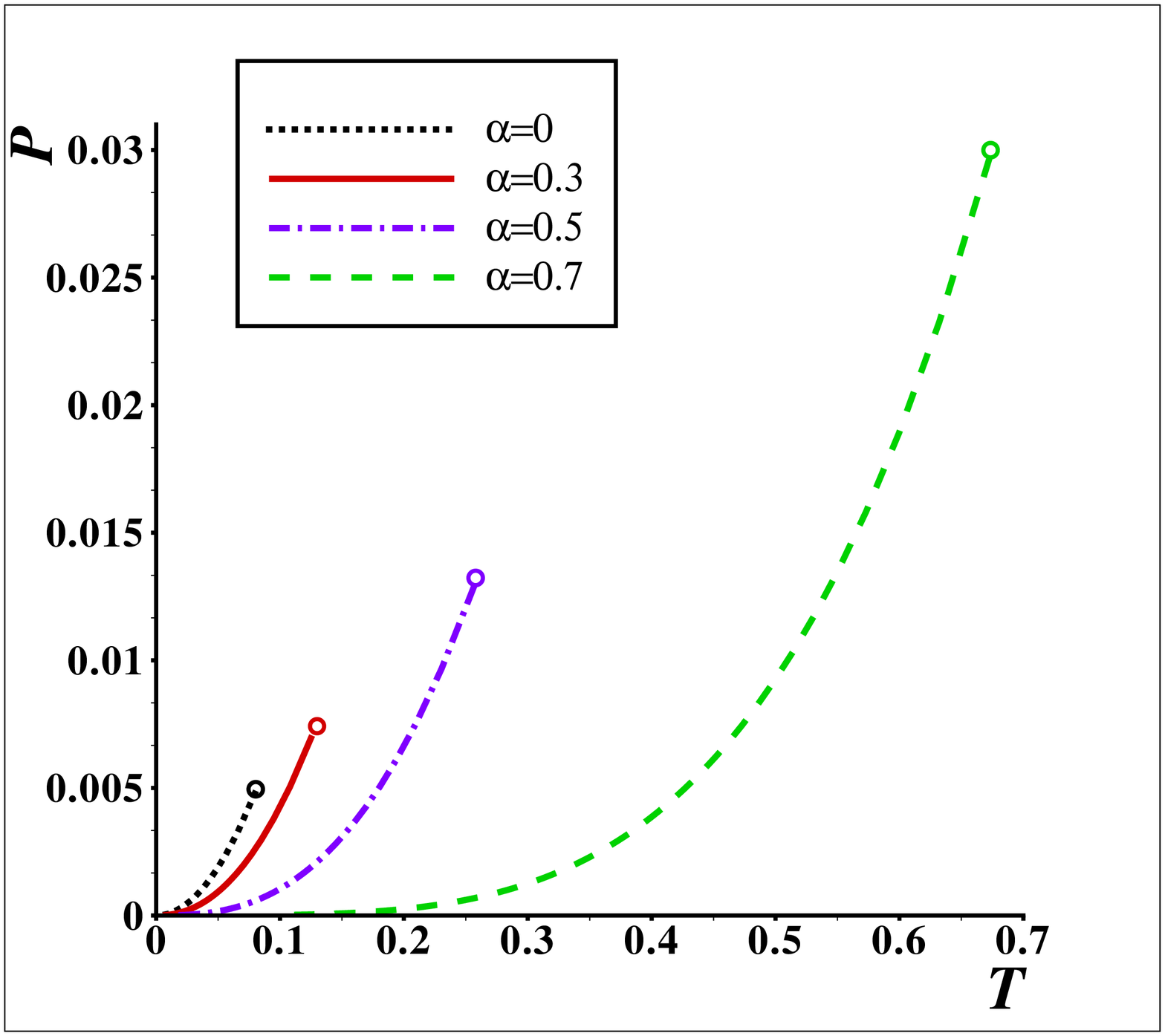}%
\end{array}%
$%
\caption{Coexistence curves of small-large black hole phase transition of
charged dilatonic black hole for $q=b=1.0$ and different values of $\protect%
\alpha $ with $n=3 $ (left) and $n=4 $ (right) in the P-T plane. The
critical points are highlighted by a small circle at the end of the
coexistence curves.}
\label{PT fig}
\end{figure}

Thermodynamic behavior of a system may be described by its thermodynamic
potential, which is the free energy in canonical ensemble. But, since we are
considering an extended phase space with variable cosmological constant, we
associate it with the Gibbs free energy $G=M-TS$ \cite{5}. It is a matter of
calculations to show that the the Gibbs free energy reduces to

\begin{eqnarray}
G=G\left( T,P\right) &=&\Big\{ \frac{k(n-2)(n-2+\alpha ^{2})b^{(n-3)\gamma }%
}{16\pi (1+\alpha ^{2})r_{+}^{(1-n)(1-\gamma )+(1-2\gamma )}}+\frac{P(\alpha
^{2}-1)(\alpha ^{2}+1)b^{(n-1)\gamma }}{(n-1)(n+\alpha ^{2})r_{+}^{n(\gamma
-1)-\gamma }}  \notag \\
&&+\frac{q^{2}(2n-3+\alpha ^{2})(\alpha ^{2}+1)b^{(3-n)\gamma }}{8\pi
(n-2+\alpha ^{2})(n-1)r_{+}^{(n-2)(1-\gamma )+\gamma }} \Big\}\omega _{n-1},
\end{eqnarray}%
where $r_{+}$ should be understood as a function of pressure and temperature
via the equation of state (\ref{eq of state}). The behaviour of the Gibbs
free energy is depicted in Fig. \ref{Gibbs fig}. This figure demonstrates
the characteristic swallowtail behavior and therefore there is a first
order phase transition in the system. Using this fact that the Gibbs free
energy, temperature and the pressure of the system are constant during the
phase transition, one can plot coexistence curves of large ($r_{+}=r_{l}$)
and small ($r_{+}=r_{s}$) charged dilatonic black holes. These are shown in
Fig. \ref{PT fig}.


\subsection{Critical exponents}

The critical exponent characterizes the behavior of physical quantities in the
vicinity of the critical point. So, following the approach of \cite{9}, we
calculate the critical exponents $\alpha ^{\prime }$, $\beta ^{\prime }$, $%
\gamma ^{\prime }$ and $\delta ^{\prime }$ for the phase transition of an $%
(n+1)$-dimensional charged dilatonic black hole. To calculate the critical
exponent\ $\alpha ^{\prime }$, we consider the entropy $S$ ( \ref{entropy})
as a function of $T$ and $V$.\ Using (\ref{volume}) we have
\begin{equation}
S=S\left( T,V\right) =\frac{\left( b^{(n-1)\gamma }\omega _{n-1}\right)
^{(1+\alpha ^{2})/(n+\alpha ^{2})}}{4}\left( \frac{n+\alpha ^{2}}{1+\alpha
^{2}}V\right) ^{(n-1)/(n++\alpha ^{2})}\ ,
\end{equation}%
which is independent of $T$. Since the exponent $\alpha ^{\prime }$ governs
the behavior of the specific heat at constant volume $C_{V}\varpropto
\left\vert t\right\vert ^{\alpha ^{\prime }}$ and
\begin{equation*}
C_{V}=T\frac{\partial S}{\partial T}\Big|_{V}=0
\end{equation*}%
one finds that $\alpha ^{\prime }=0$.

To obtain the other exponents, we define the reduced thermodynamic variables
\begin{equation*}
p\equiv \frac{P}{P_{c}},\quad \nu \equiv \frac{v}{v_{c}},\quad \tau \equiv
\frac{T}{T_{c}}.
\end{equation*}%
So, equation of state (\ref{PvT}) translates into `the law of corresponding
state ':

\begin{eqnarray}
p &=&\frac{2(2n-2+2\alpha ^{2})}{(2n-3+\alpha ^{2})(1-\alpha ^{2})}\frac{%
\tau }{\nu }+\frac{(2n-2+2\alpha ^{2})(1+\alpha ^{2})}{(\alpha
^{2}-1)(2n-4+2\alpha ^{2})}\frac{1}{\nu ^{2-2\gamma }}  \notag \\
&&+\frac{(1+\alpha ^{2})}{(2n-3+\alpha ^{2})(n-2+\alpha ^{2})}\frac{1}{\nu
^{\gamma +1+(2n-3)(1-\gamma )}}.  \label{law}
\end{eqnarray}%
\newline
In order to find the other critical exponents, we follow the method of Ref.\
\cite{9} and expand Eq.\ (\ref{law}) near the critical point
\begin{equation}
t=\tau -1,\quad \omega =\nu ^{\varepsilon }-1=\dfrac{V}{V_{c}}-1.  \label{29}
\end{equation}%
One obtains
\begin{equation}
p=1+At-Bt\omega -C\omega ^{3}+O\left( t\omega ^{2},\omega ^{4}\right) ,
\label{ptw}
\end{equation}%
where
\begin{eqnarray}
A &=&\frac{1}{\rho _{c}}=\frac{4(n-1+\alpha ^{2})}{(2n-3+\alpha
^{2})(1-\alpha ^{2})},\quad  \\
B &=&\frac{1}{\varepsilon \rho _{c}}=\frac{4(n-1+\alpha ^{2})(1+\alpha ^{2})%
}{(2n-3+\alpha ^{2})(1-\alpha ^{2})(n+\alpha ^{2})}, \\
C &=&\frac{2(n-1+\alpha ^{2})}{3(1+\alpha ^{2})^{2}\varepsilon ^{3}}=\frac{%
2(n-1+\alpha ^{2})(1+\alpha ^{2})}{3(n+\alpha ^{2})^{3}}.
\end{eqnarray}%
Denoting the volume of small and large black holes by $\omega _{s}$ and $%
\omega _{l}$, respectively, differentiating Eq.\ (\ref{ptw}) with respect to
$\omega $\ at a fixed $t<0$, and applying the Maxwell's equal area law\ \cite%
{8} one obtains
\begin{eqnarray}
p &=&1+At-Bt\omega _{l}-C\omega _{l}^{3}=1+At-Bt\omega _{s}-C\omega _{s}^{3}
\notag \\
0 &=&-P_{c}\int_{\omega _{l}}^{\omega _{s}}\omega \left( Bt+3C\omega
^{2}\right) d\omega ,  \label{EqualA}
\end{eqnarray}%
Equation (\ref{EqualA}) leads to the unique nontrivial solution
\begin{equation}
\omega _{l}=-\omega _{s}=\sqrt{-\frac{Bt}{C}},  \label{oml}
\end{equation}%
which  gives the order parameter $\eta =V_{c}\left( \omega _{l}-\omega
_{s}\right) $ as
\begin{equation}
\eta =2V_{c}\omega _{l}=2\sqrt{-\frac{B}{C}}t^{1/2}.
\end{equation}%
Thus, the exponent $\beta ^{\prime }$ which describes the behaviour of the
order parameter $\eta $ near the critical point is $\beta ^{\prime }=1/2.$
To calculate the exponent $\gamma ^{\prime }$, we may determine the behavior
of the isothermal compressibility near the critical point
\begin{equation*}
\kappa _{T}=-\frac{1}{V}\frac{\partial V}{\partial P}\Big|_{T}\varpropto
\left\vert t\right\vert ^{-\gamma ^{\prime }}.
\end{equation*}%
Since $dV/d\omega =V_{c}$, the isothermal compressibility near the critical
point reduces to
\begin{equation}
\kappa _{T}=-\frac{1}{V}\frac{\partial V}{\partial P}\Big|_{T}\varpropto
\frac{V_{c}}{BP_{c}t},
\end{equation}%
which shows that $\gamma ^{\prime }=1$. Finally the "shape" of the critical
isotherm\ $t=0$ is given by (\ref{ptw})
\begin{equation}
p-1=-C\omega ^{3},
\end{equation}%
which indicates that $\delta ^{\prime }=3$.

Although the reduced pressure $p$ in Eq. (\ref{law}) depends on the dilaton
parameter $\alpha $ and the dimension of the spacetime, the critical
exponents associated with the charged dilatonic black hole in $(n+1)$
dimensions are independent of them. This is consistent with the belief that
the critical exponents are universal and do not depend on the details of the
physical system.

\section{Summary and Conclusions\label{sec3}}

In this paper, we investigated the critical behavior of charged dilatonic
black holes of Einstein-Maxwell-dilaton gravity in the presence of the
potential $V(\Phi )$ given in Eq. (\ref{VPhi}) while $\Lambda $ treated as a
thermodynamic quantity that can vary. Using the known thermodynamic
quantities, we identified the thermodynamic pressure and volume of the
system and obtained the Smarr-type relation for charged dilatonic black
holes. As in the cases of rotating \cite{3}, Taub-Nut and Taub-bolt black
holes \cite{Nut}, we found that the volume is different from the
geometrical volume. In canonical ensemble we wrote the equation of state as $%
P=P(v,T)$ and plot $P-v$ isotherm diagrams. These diagrams are similar to
those of Van der Waals fluid and indicate a first-order phase transition
between small and large black holes in temperature bellow critical
temperature. We found that for any nontrivial value of the charge, there
exists a critical temperature $T_{c}$ below which this phase transition
occurs. Moreover, we studied the behavior of certain physical quantities
near the critical point and calculated the pressure, temperature and volume
at the critical temperature. Although Eq. (\ref{P1}) has solution for the
case of $k=-1$ for $\alpha >1$, critical behavior do not occur in this case.
This is due to the fact that the critical ratio $\rho _{c}=P_{c}v_{c}/T_{c}$
becomes negative for $\alpha >1$. Thus, we limited ourselves to the case of $%
k=1$ and $\alpha <1$ for which the critical ratio is positive. The
characteristic swallowtail behavior of the Gibbs free energy was another
indication of the first-order phase transition in the system. Using this
fact that the Gibbs free energy, temperature, and pressure of the system
are constant during the phase transition, we plotted the coexistence curves
of large-small charged dilatonic black holes and again we showed that a
first-order phase transition in temperature bellow critical temperature
occurs. Finally, we calculated the critical exponents and found that they
are universal and are independent of the details of the system although the
reduced pressure $p$ in Eq. (\ref{law}) depends on the dilaton parameter $%
\alpha$ and the dimension of the spacetime.

\acknowledgments{We thank from the Research Council of Shiraz
University. This work has been supported financially by Research
Institute for Astronomy \& Astrophysics of Maragha (RIAAM), Iran.}

\end{document}